\documentclass[aps,preprint]{revtex4}%
\usepackage{amsfonts}
\usepackage{amsmath}
\usepackage{amssymb}
\usepackage{graphicx}%
\providecommand{\U}[1]{\protect\rule{.1in}{.1in}}

\begin{document}
\preprint{ }
\title[Short title for running header]{CLASSICAL ('ONTOLOGICAL') DUAL STATES IN QUANTUM THEORY AND THE MINIMAL GROUP REPRESENTATION HILBERT SPACE}
\author{Diego J. CIRILO-LOMBARDO}
\affiliation{M. V. Keldysh Institute of the Russian Academy of Sciences, Federal Research
Center-Institute of Applied Mathematics, Miusskaya sq. 4, 125047 Moscow,
Russian Federation}
\affiliation{CONICET-Universidad de Buenos Aires, Departamento de Fisica, Instituto de
Fisica Interdisciplinaria y Aplicada (INFINA), Buenos Aires, Argentina.}
\author{Norma G. SANCHEZ}
\affiliation{The International School of Astrophysics Daniel Chalonge - Hector de Vega, CNRS,
INSU-Institut National des Sciences de l'Univers, Sorbonne University  75014
Paris, France.\\
Norma.Sanchez@obspm.fr \\
https://chalonge-devega.fr/sanchez
}
\;
\date{ \today}
\pacs{PACS number}

\begin{abstract}
We investigate the classical aspects of Quantum theory and under which description Quantum theory does {\it appear Classical}. Although such descriptions or variables are known as "ontological" or "hidden", they are {\it not hidden} at all, but are {\it dual classical} states (in the sense of the general classical-quantum duality of Nature). We analyze and interpret the 
dynamical scenario in an inherent quantum structure: {\bf(i)} We show that the use of the known $\left\vert \varphi\,\right \rangle $  states in the circle (London 1926, 't Hooft 2024),  takes a
true dimension  {\it only} when the system is subjected to the 
{\it minimal group representation} action of the 
metaplectic group $Mp(n)$. $Mp(n)$ Hermitian structure fully covers the symplectic  $Sp(n)$ group and in certain cases $OSp(n)$.  
{\bf(ii)} We compare the  
 circle $\left\vert \varphi\,\right \rangle $ states and the  
the cilinder \, $ \left\vert \xi\,\right \rangle $ states {\it in configuration space} with the two sectors of the full $Mp(2)$ Hilbert space corresponding to the {\it even} and {\it odd}  $n$ harmonic oscillators and their {\it total} sum. {\bf(iii)} We compute the projections of the $Mp(2)$ 
states on the circle $\left\langle \,\varphi\,\right\vert $ and on the  cilinder $\left\langle \,\xi\,\right\vert $ states. The known London circle states are {\it not normalizable}. We compute 
here the {\it general} coset coherent states $\left\langle \,\alpha\,, \varphi \right\vert  $ in the circle,  $\alpha$ being the coherent complex parameter: It allows fully {\it  normalizability} of the complete set of the circle states. {\bf(iv)} The London states (ontological in t'Hooft's description) { \it completely classicalize }  the inherent quantum structure {\it only} under the action of the $Mp(n)$  {\it Minimal Group Representation }.
{\bf(v)}  For the coherent states in the cylinder (configuration space), all functions are analytic in the disc $\left \vert \,z \,=\, \omega\, e^{-i\varphi} \right \vert \, < \,1 $.  For the general coset coherent states $\left \vert\alpha, \varphi \right \rangle$ in the circle, the complex variable is $ z^{\prime}  \;=\;z\; e^{\,-i\,\alpha^{\ast}%
/2}$\,:  the analytic function  is modified by the complex phase
$(\,\varphi-\alpha^{\ast}/2\,)$.
{\bf(vi)} The analiticity  $\left\vert \,z^{\prime}\right\vert \,= \,\left\vert z \right\vert e^{-\operatorname{Im}\alpha/2}\,<\,1 $ occurs with $ \operatorname{Im}\alpha \;\neq \;0 $ because of normalizability,  and 
 $ \operatorname{Im}\alpha > 0 $ because of  the Identity condition. 
 The circle topology induced by the 
$\left\langle \alpha,\varphi\right\vert $ coset coherent state, also modifies the ratio of the disc due the displacement 
by the coset. {\bf(vii)} For the coset coherent cilinder states {\it in configuration space}, the classicalization is stronger due to screening exponential factors $e^{-2\,n^{2}}$,  $e^{-(\,2n+1/2\,)}$ and $e^{-(\,2n+1/2\,)^{2}}$ for large $n$ arising in the $Mp(2)$ projections on them. The generalized Wigner function shows a bell-shaped distribution and {\it stronger classicalization}  than the square norm functions. The application of the Minimal Group Representation immediately classicalizes the system, $Mp(2)$ emerging as the group of the {\it classical-quantum duality} symmetry.
\end{abstract}
\volumeyear{year}
\volumenumber{number}
\issuenumber{number}
\eid{identifier}
\date{\today}

\maketitle
\maketitle
\tableofcontents

\section{Introduction and Results}

Nature is Quantum. Classical states are contained in a Quantum description. Quantum theory is in full development, whatever be in its own concepts and extensions, interpretations, and yet to be understood effects and manifestations, or in new applications, Artificial Intelligence, Quantum computing, technological and fundamental experimental searches. See for example Refs \cite {Nobel-2022}, \cite{MacLoughlin-Sanchez}, \cite{Nobel-2024} and Refs therein. Is not the purpose of this Introduction to review it here.   

\bigskip

{\bf In this paper}, without entering in all the issues of the interpretations of Quantum theory,  we consider  the subject of the Classical aspects of Quantum theory and under which description Quantum theory does {\it appear Classical}. \\

Usually, the variables providing such a description are known as "ontological" or  ‘local
 hidden variables’, - although {\it they are not hidden at all}, and we agree with 't Hooft  Ref \cite{tHooft1} who recently considered these variables as {\it not hidden}. \\ 
 
 Usually, such classical or "ontological" variables are assumed to describe the results of the real measurements performed on a given quantum system,  (ontological meaning here essential or existential, conceptual or substantial, for an entity or representation).
 
 \bigskip
 
The  general Classical- Quantum (wave-particle, de Broglie) duality  is one deep foundational property of quantum theory and does remain a crucial one. More recently , Refs. \cite{NSPRD2021}\cite{NSPRD2023}\cite{Sanchez2019a}\cite{Sanchez2019b}\cite{Sanchez2019c} this concept have been extended to include  
gravity at the Planck scale and beyond it (classical-quantum gravity duality), which is general, and does not depend on the number or type of space-time dimensions, nor manifolds  (compactified or not), or on other considerations.

\bigskip

The Classical "ontological or existing" states are in fact {\it Classical Dual states} and we do prefer this last term because it is physically precise, appropriate and meaningful for these states.

\medskip

 The time evolution of the quantum harmonic oscillator does appear identical to the classical motion: a rigid rotating motion in phase space with the oscillator frequency. The quantum states of light are harmonic oscillators and have the same time evolution.

\bigskip

We also consider the {\it Wigner function} which is a useful tool for the comparison of the classical and quantum dynamics of these states in phase space, and to analyze the classical behavior in general. It allows to obtain the probability distribution for these states in phase space. The Wigner or quasiprobability distribution, namely $ W(q,p)$ is symmetric in the reflection (time and space) symmetries, e.g $ W (- q, - p) =  W (q, p)$. 

\bigskip

{\bf In this paper}:
\begin{itemize}

\item{In order to interpret the 
dynamical scenario connected with an inherent quantum structure, we show that the use of the known   states in the circle ('t Hooft 2024 \cite{tHooft1}, London 1926 \cite{london}),  takes a
true dimension only when the system is subjected to the 
minimal group representation under the action of the
metaplectic group $Mp(n)$. Let us recall that $Mp(n)$ fully covers the symplectic group $Sp(n)$ and in
certain cases its Hermitian structure can be extended to $OSp(n)$.}

\item{ We consider the Metaplectic group $Mp(2n)$, e.g the double covering of the symplectic group $ Sp(2n)$.  
{\it Discretization} arises naturally and directly from the basic states of the metaplectic representations with an interesting feature to be
highlighted here: the decomposition of the $SO(2,1)$ group into its two
irreducible representations encompasses the span of the \textit{even} $\left\vert
\;2n\;\right\rangle $ and \textit{odd} $\left\vert \;2n\;+\;1\;\right\rangle $ 
states, respectively, ($n = 1,\,2,\,3\,...$ ) of the harmonic oscillator, whose entirety is covered by the
metaplectic group. For $n \rightarrow \infty$, the spectrum becomes continuum as it must be.}
 
\item{ In the metaplectic representation,  the general or complete
states must be the sum of the two types states: {\it even and \it odd $n$} states, 
because they span respectively the two sectors of the Hilbert space, $\mathcal{H}_{1/4}$ and
$\mathcal{H}_{3/4}$,  completely covered by the Metaplectic symmetry group $M_{p}(2)$\,: $\mathcal{H}_{1/4}%
\oplus\mathcal{H}_{3/4}$. Based on the highest eigenvalue of the number operator occurring in the complete Hilbert space, the two unitary irreducible representations of $Mp(2)$ are denoted  $\mathcal{H}_{1/4}$ \,(even states)  and $\mathcal{H}_{3/4}$ \,(odd states).}

\item{ We compare the complete and fully normalizable $Mp(2)$ states with the $\left\vert \varphi\,\right \rangle $ states in the circle (London states, phase space, and recently considered by 't Hooft).}

\item {We also consider general coherent states \, $ \left\vert \xi\,\right \rangle $ in 
the cilinder {\it in configuration space} and compare them with the sectors of the $Mp(2)$ Hilbert space,
and with the {\it total} (sum of the two sectors) states $(+)$ and $(-)$, corresponding to the even $(2n)$ and odd $(2n+1)$ states of the harmonic oscillators. 
We compute the projections of the Mp(2)  $\left\vert \,\Psi^{\pm} \left(  \omega\right)
\,\right\rangle$
states on the circle $\left\langle \,\varphi\,\right\vert  $ and on the  cilinder $\left\langle \,\xi\,\right\vert $ states.} 

\item {The known circle (London, t' Hooft) states are {\it coherent not normalizable states}. We compute 
here the {\it general} coherent states in the circle $\left\langle \,\alpha\, ,\varphi \,\right\vert  $ which include the complex parameter $\alpha$ characteristic of coherent states, whose meaning appears clearly allowing the fully {\it finite normalization} of the complete set of states. The power of the general {\it coset group construction procedure} of coherent states does show up here.} 
 \end{itemize}

{\bf Main implications of 
 the results of this paper are the following:} 

\bigskip

\begin{itemize}
\item {{\bf(i)} The London states (ontological in t'Hooft's description) { \it classicalize completely }  the inherent quantum structure {\it only} under the application of the Minimal Group Representation with  the $Mp(n)$ group taking the main role.}

\item{{\bf(ii)} The action of the Metaplectic group on the "ontological" (London) states
breaks the invariance under time reversal assumed for the dynamics of the
particle in the circle (arrow of time).}
\item{{\bf(iii)}  In the case of the coherent states of a particle in the cylinder (configuration space), we can also assign to them the variable $ z \,=\, \omega\, e^{-i\varphi}$ as in the case of the particle states in a circle, (London states, phase space). All functions are analytic in the disc $\left \vert \,z\right \vert \, < \,1$.  For the general coherent states in the circle  $\left \vert\alpha, \varphi \right \rangle$ the complex variable is $ z^{\prime} \, = \, \omega \,e^{\,i\,\left(\,  \varphi\,-\,\alpha^{\ast}/2\,\right)  }\;=\;z\; e^{\,-i\,\alpha^{\ast}%
/2\,}$
$:$ the analytic function in the disc is modified by the complex phase
$(\,\varphi-\alpha^{\ast}/2\,)$, $\alpha$ 
 being the characteristic coherent state complex parameter.}

\item{{\bf(iv)} The analiticity condition  on the disk : $\left\vert \,z^{\prime}\right\vert \,= \,\left\vert z \right\vert e^{-\operatorname{Im}\alpha/2}\,<\,1 $ occurs with the condition $ \operatorname{Im}\alpha \;\neq \;0 $ arising from the normalizability,  and with 
 $ \operatorname{Im}\alpha > 0 $ arising from the Identity condition. 
 The topology of the circle induced by the coset coherent state
$\left\langle \alpha,\varphi\right\vert $ not only modifies the
phase of $\omega$ (e.g: $\omega \,e^{i\,\left( \, \varphi-\alpha^{\ast}/2 \,\right)
} = z^{\prime}\,)$ but also the ratio of the disc due the displacement generated
by the action of the coset.}

\item{{\bf(v)} The norm of the projection of the $Mp(2)$ states on the cilinder and on general circle coherent states  clearly shows a very fast decreasing as $n$ increases or {\it classicalization}. In the case of the  cilinder states {\it in configuration space} the decreasing is stronger than in the circle phase space states, due to the exponentials $e^{-2\,n^{2}}$,  $e^{-(\,2n+1/2\,)}$ and $e^{-(\,2n+1/2\,)^{2}}$, arising in these configuration space projections.

The generalized Wigner function  for the circle states   displays a classicalized distribution, more bell- shaped that the square norm function of these states.\\

More remarks are presented in the Conclusions.} 
\end{itemize}

{\bf This paper is organized as follows:}
In Section II we describe the 
quantum dynamics on the circle and its phase space states (London states). 
In Section III we summarize the Metaplectic Minimal Group Representation approach, its group content, double covering and fully complete Hilbert space of states. In Section IV both the $Mp(2)$ basic states and the ontological states in the circle are compared together with their their mutual scalar products, which shows how classicalization does occurs in this case. In Section V we consider the general coherent states in a cilinder in configuration space and compute the projections of the $Mp(2)$ total states on them, which show a still more strong classicalization with respect to the London (circle, phase space) states.  Section VI discusses the implications of the $Mp(2)$ Representation Reduction on theses states and the analysis and results of the previous sections. Sections VII and VIII the general normalized coset coherent states 
on the circle and the projected  $Mp(2)$  Reduction on them are computed and analysed. Section IX summarizes the Conclusions.

\section{"Ontological" States and the Minimal Group Representation}

As we showed earlier Ref \cite{universe} , there is an even more general principle in the
fundamental structure of quantum spacetime: {\it the principle of minimal group
representation}, which allows us to obtain, consistently and simultaneously, a
natural description of the dynamics of spacetime and the physical states
admissible within it.

\medskip

The theoretical design is based on physical states which
are the mean values {}{}of the metaplectic group generators $Mp(n)$, the double
covering $SL(2C)$ in vector representation, relative to the coherent states bearing the spin weight. 

\medskip

In this theoretical context, there is a connection
between the dynamics given by the symmetry generators of the metaplectic group
and the physical states (mappings of the generators through bilinear
combinations of the ground states), Refs \cite{universe}, \cite{cirilo-sanchezPRD}.

\medskip

Let us now see how to apply this principle to
the problem of the construction of classical variables in quantum theory as considered by t' Hooft in Ref \cite{tHooft1}. Therefore, we first need to consider the quantum dynamics of a particle on the circle, which we describe in the next Section.

\subsection{II Theoretical aspects of the quantum dynamics on the circle}

{\bf(i)} The starting point to have into account to describe a free particle on a circle is its Hamiltonian %
\[
H\; =\;L = \;\frac{1}{2}\overset{\cdot}{\varphi}^{2}(t)\;\;\;\left(e.g:\;\;\overset{\cdot}%
{\varphi}\;=\; J\right)
\]
here with unit mass and velocity,  $\varphi$ being the angle position, with  
period $2 \pi$:
$$  \varphi\,(t) =  \varphi (0) + t $$ 
Classically,%
\[
\left\{  \varphi,J\right\} \; =\;1
\]
Quantically,  (operator level):%
\begin{equation}
\left[  \widehat{\varphi},\widehat{J}\right]  \;=\; i \label{c}%
\end{equation}
The best candidate for the position operator (well defined in Hilbert space) is: 
\[
U\;=\;e^{\,i\,\widehat{\varphi}} \;\;\;\text{
\ \ \ \ (\,U is unitary\,)}
\]
And it is easy to see from the above equations that:%
\begin{equation}
\left[  \widehat{J}\;,\;U\right] \; = \;U \label{a}%
\end{equation}

{\bf (ii)} Let now consider the eigenstates%
\begin{equation}
\widehat{J}\;\left\vert \;j \;\right\rangle \;= \;j\;\left\vert \;j\;\right\rangle \label{b}%
\end{equation}
From Eqs (\ref{a}) and (\ref{b}) we have:%
\begin{align*}
U\;\left\vert \;j\;\right\rangle  &  \;=\;j\left\vert \;j+1\;\right\rangle ,\\
U^{\dagger}\;\left\vert\; j\;\right\rangle  &  =\;j\left\vert\; j-1\;\right\rangle
\end{align*}
e.g. $U^{\dagger}$ and $U$ are ladder operators. 

Now, it is esily seen the additional
properties the states $\left\vert \;j\;\right\rangle $ satisfy:%
\[
\left\langle \;i\;\right\vert \left\vert \;j\;\right\rangle \;=\;\delta_{i\,j}\text{
\ \ \ \ (orthogonality)}%
\]%
\[
\underset{j}{%
{\displaystyle\sum}
}\left\vert \;j\;\right\rangle \left\langle \;j\;\right\vert \;=\;\mathbb{I}\text{
\ \ \ \ \ (completeness)}%
\]
As we see, we have all the ingredients to implement the principle of minimal
representation of the group: 

- (i) the basis $\left\vert \;j\;\right\rangle
$\textit{ isomorphic to } the basis $\left\vert \;n\;\right\rangle$ of the harmonic oscillator,\\
- (ii) a \textit{symplectic structure},
and (iii) the conmutator Eq.(\ref{c}) that
allows to build the two operators: $$a\;=\;\frac{1}{\sqrt{2}}\left(  \widehat{\varphi
}\;+\;i\widehat{J}\right) \;\;\; \text{and} \;\;\; a^{+}=\;\frac{1}{\sqrt{2}}\left(  \widehat
{\varphi}\;-\;i\widehat{J}\right)$$
\\
that is, $a$ and $a^{+}$ are ladder operators satisfying:
\begin{equation}
    \left[ a \;, \;a ^+    \right ]\; = \; 1 
\end{equation}

Let us now consider the Metaplectic group approach. 

\section{The Metaplectic Minimal Group Representation Approach}

In Ref \cite{universe}, \cite{cirilo-sanchezPRD} and refs. therein, a group-theoretic approach was developed to obtain
the metric (line element) as the central geometrical object associated to a
discrete quantum structure of the spacetime for a quantum theory of gravity.

In summary for the purpose of this paper, the main characteristics of this framework  are the following: 
\begin{itemize}
\item{ Such
an emergent metric is obtained from a Riemmanian superspace and is described as
a physical coherent state of the underlying cover of the $SL(2C)$ group:
Interestingly, it seems necessary to consider the {\it full cover of the symplectic
group}, which is the metaplectic group $Mp(n)$, its spectrum for all $n$ leads in
particular for very large $n$ to continuous spacetime.}

\item{The main and fundamental importance of this quantum description is based on
the phase space of a relativistic particle in the \textit{ natural} superspace of bosonic and fermionic coordinates that allow preserving at the quantum level
the square root forms of geometric operators (e.g.  the Hamiltonian or the Lagrangian).} 

\item{Such natural characteristic of this description is ensured by a
complete bosonic realization of the underlying algebra through creation
and annihilation operators of the harmonic oscillator that establish the
gradation of it.} 
\item{ The {\it discrete} structure of the spacetime arises directly from the basic
states of the metaplectic representation with an interesting feature to be
highlighted here: the decomposition of the $SO(2,1)$ group into its two
irreducible representations encompasses the span of \textit{even} $\left\vert
\;2n\;\right\rangle $ and \textit{odd} $\left\vert \;2n\;+\;1\;\right\rangle $
states, ($n = 1,\,2,\,3\,...$ ), respectively, whose entirety is covered by the
metaplectic group.}

\item{ In the metaplectic representation,  the general or complete
states must be the sum of the two types of states: the even and odd $n$ states, 
because they span respectively the two sectors of the Hilbert space, $\mathcal{H}_{1/4}$ and
$\mathcal{H}_{3/4}$, whose complete covering is $\mathcal{H}_{1/4}%
\oplus\mathcal{H}_{3/4}$ corresponding to the Metaplectic symmetry group $M_{p}(2)$.} 

\item{This yields the relativistic quantum metric of the
discrete structure spacetime as the fundamental basis for a quantum theory of
gravity. For increasing numbers of levels $n$, the metric solution goes to the
continuum and to the classical general relativistic manifold as it should be.}

\item{The double covering of {\it even and odd} $n$ states and {\it their sum} in order to have the
complete Hilbert space reflects here the CPT completeness of the theory, and such
property is the reflection of unitarity. As we know, the metaplectic
group $M_{p}(2)$ acts irreducibly on each of the subspaces $\mathcal{H}_{1/4},$
$\mathcal{H}_{3/4}$ (even and odd sectors) by which the total Hilbert space (i.e.,$\mathcal{H}$) is divided, according to the $M_{p}(2)$ Casimir operator which gives precisely the values $k\;=\;1/4,\;\,3/4$:}
\end{itemize}
\[
K^{2}\;=\;K_{3}^{2}\;-\;K_{1}^{2}\;-\;K_{2}^{2}\;=\;k\;\left(
\,k\,-\,1\,\right)  \;=\;-\,\frac{3}{16}\;\mathbb{I}%
\]
Therefore,
\begin{equation}
\mathcal{H}_{1/4}\;=\;\text{Span}\left\{  \;\left\vert \;n\;\text{{\it even}}%
\;\right\rangle \;\text{states}:\;n\;=\;0,\,2,\,4,\,6,\,...\;\right\}
\end{equation}%
\begin{equation}
\mathcal{H}_{3/4}\;=\;\text{Span}\left\{  \;\left\vert \;n\;\text{{\it odd}}%
\;\right\rangle \;\text{states}:\;n\;=\;1,\,3,\,5,\,7,\,....\;\right\}
\end{equation}

Based on the highest eigenvalue of the number operator occurring in the complete\\ $\mathcal{H} \;\equiv
\;\mathcal{H}_{1/4}\oplus\mathcal{H}_{3/4}$:
$$T_{3}\,\left\vert \,
n\,\right\rangle \, = \,-\,\frac{1}{2}\left(  \,n\,+ \,\frac{1}{2}\,\right)
\left\vert \,n\,\right\rangle ,
$$ 
the two unitary irreducible
representations (UIR) of $Mp(2)$ are denoted as:
\begin{equation}
(UIR) \quad\text{ restricted \;to \;}\mathcal{H}_{1/4} \quad\rightarrow
\quad\mathcal{D}_{1/4} \; \in\; \; Mp(2)\\
\end{equation}
\begin{equation}
(UIR)\quad\text{ restricted \;to \;}\;\mathcal{H}_{3/4} \;\rightarrow
\quad\mathcal{D}_{3/4} \quad\in\; Mp(2)
\end{equation}
\;
It is notable that in the general case, \thinspace$Sp(2m)$\thinspace\ can be
embedded somehow in a larger algebra as \thinspace$(Sp(2m)\,+\,R^{2m})%
$\ admitting an Hermitian structure with respect to which it becomes the
orthosymplectic superalgebra \thinspace\ $Osp(2m,1)$. 

\bigskip

Consequently, the
metaplectic representation of \thinspace$Sp(2m)$\thinspace\ extends to an
irreducible representation (IR) of \thinspace$Osp(2m,1)$\thinspace\ which can
be realized in terms of the space $H_{h}$ of all holomorphic functions
\thinspace$h$:
$$C^{m}\;\rightarrow\;C/\int\left\vert h\left(  z\right)
\right\vert ^{2}e^{-\left\vert z\right\vert ^{2}}\,d\lambda\left(  z\right)
\,<\infty$$
with $\lambda\left(  z\right)  $ the Lebesgue measure on $C^{m}$. 

\bigskip

The restriction of the $Mp\left(n\right)  $ representation to $Sp(2m)$%
,\thinspace\ implies that the two irreducible (even and odd $n$) sectors are supported by the
subspaces $H^{\pm}_h$ of the holomorphic functions space $H_h$.  $H^{+}_h$ and $H^{-}_h$ are the (closed) spans 
of the set of functions: $$z^{n}\;\equiv\;\left(  z_{1}^{n_{1}},....,\,z_{m}^{n_{m}%
}\right) $$ 
where \thinspace\ $n_{\theta}\,\in\,Z,$\;\; $\left\vert n\right\vert
\;=\;\sum n_{\theta}$, \textit{even}  and \textit{odd}, ($H^{+}_h$ and $H^{-}_h$), respectively.

\medskip

\subsection
{\textbf{III Mp(2), SU(1,1) and Sp(2)}}

All the groups $Mp(2)$, $Sp(2,R)$, and $SU(1,1)$ are three dimensional. It
is possible to parameterize them in several ways that make the homomorphic
relations particularly simple. We use two of such parameterizations, in terms of the $Mp(2)$ group generators $(T_1, T_2, T_3)$ and the angles $(\alpha_1, \alpha_2, \alpha_3)$,  both of
which are described as:
\[
Mp\,(2)\;\;\rightarrow\;\;e^{-\,i\, \alpha_{1}T_{1}},\;\;e^{-\,i\, \alpha_{2}T_{2}%
},\;\; e^{-\,i \,\alpha_{3}T_{3}}%
\]
\vspace{-0.5em}
\[
Sp\,(2 {R}) \;\rightarrow\;\left(
\begin{array}
[c]{cc}%
e^{\frac{1}{2}\alpha_{1}} & 0\\
0 & e^{-\frac{1}{2}\alpha_{1}}%
\end{array}
\right)  ,\left(
\begin{array}
[c]{cc}%
\cosh\frac{1}{2}\alpha_{2} & \sinh\frac{1}{2}\alpha_{2}\\
\sinh\frac{1}{2}\alpha_{2} & \cosh\frac{1}{2}\alpha_{2}%
\end{array}
\right)  ,\left(
\begin{array}
[c]{cc}%
\cos\frac{1}{2}\alpha_{3} & - \sin\frac{1}{2}\alpha_{3}\\
\sin\frac{1}{2}\alpha_{3} & \cos\frac{1}{2}\alpha_{3}%
\end{array}
\right)
\]

\[
SU(1,1)\rightarrow\left(
\begin{array}
[c]{cc}%
\cosh\frac{1}{2}\alpha_{1} & \sinh\frac{1}{2}\alpha_{1}\\
\sinh\frac{1}{2}\alpha_{1} & \cosh\frac{1}{2}\alpha_{1}%
\end{array}
\right)  ,\left(
\begin{array}
[c]{cc}%
\cosh\frac{1}{2}\alpha_{2} & i\sinh\frac{1}{2}\alpha_{2}\\
-i\sinh\frac{1}{2}\alpha_{2} & \cosh\frac{1}{2}\alpha_{2}%
\end{array}
\right)  ,\left(
\begin{array}
[c]{cc}%
e^{\frac{i}{2}\alpha_{3}} & 0\\
0 & e^{-\frac{i}{2}\alpha_{3}}%
\end{array}
\right)
\]

\bigskip

where the angle $\alpha_{3}$ has the range $(-4\pi,4\pi]$ for $Mp(2)$,
and the range $(-2\pi,2\pi]$ for $Sp\,(2,R)$ and $SU(1,1)$.

\bigskip

Let us consider the brief description of the relevant symmetry group to
perform the realization of the Hamiltonian operator of the problem. This group
specifically is the Metaplectic $Mp\left(  2\right)  $ as well as the groups
that are topologically covered by it. The generators of $Mp\left(  2\right)  $
are the following :
\begin{align}
T_{1}  &  \;\;= \;\;\frac{1}{4}\;\left(  \, q\,p\;+\;p\,q\;\right)  \;\; =
\;\;\frac{i}{4}\;\left(  \, a^{+2}-\;a^{2}\,\right)  ,\label{cr}\\
T_{2}  &  \;\;= \;\;\frac{1}{4}\;\left(  \,p^{2}\;-\;q^{2}\,\right)  \;\;=
\;-\;\frac{1}{4}\;\left(  \, a^{+2} + \;a^{2}\right)  ,\nonumber\\
T_{3}  &  \; \;= \;-\;\frac{1}{4} \;\left(  \, p^{2}\;+\;q^{2}\,\right)  \;\;
= \;-\;\frac{1}{4}\;\left(  a^{+}a \; + \;a\,a^{+} \,\right) \nonumber
\end{align}
with the commutation relations,
\[
\left[  \; T_{1}, \;T_{2}%
\;\right]  \; = \;-\;i\;T_{3}%
\]
\[
\left[  \; T_{3},\;T_{1}\;\right]  \;=\;i\;T_{2},\;\;\;\;\;\left[  \; T_{3},
\;T_{2}\;\right]  \; =\;-\;i\;T_{1},
\]

being $(q,\,p)$, or alternatively $(a,\,a^{+})$, the variables of the standard
harmonic oscillator, as usual.

\medskip

If we rewrite the commutation relations as:
\[
\left[  \; T_{3}\;,\;T_{1}\;\pm\;i\;T_{2}\;\right]  \; = \;\pm\;\left(
\;T_{1} \;\pm\; i\;T_{2} \;\right),
\]
\[
\left[  \; T_{1} \;+ \;i_{\;}%
T_{2}\;, \;T_{1}\;-\;i\;T_{2}\;\right]  \; = \;- \; 2\;T_{3},%
\]
\ 

we see that the states $\left\vert \;n\;\right\rangle $ are eigenstates of
$T_{3}$:%
\[
T_{3}\;\left\vert \;n\;\right\rangle \;=\;-\;\frac{1}{2}\;\left(  n+\frac{1}%
{2}\right)  \left\vert \;n\;\right\rangle
\]
And it is easy to see that:\
\[
T_{1}\;\;+\;\;i\,T_{2}\;=\;-\;\frac{i}{2}\;a^{2},\;\;\;\;\;\;\;T_{1}
-\;i\;T_{2}\;\;=\;\;\frac{i}{2}\;a^{+2}.
\]

\section{ The Mp(2) Basic States vs. Ontological States in the Circle}

Let us look at the sectors $s = 1/4$ and $s = 3/4$ of the Hilbert space spanned by the $Mp(2)$ coherent 
states $\left\vert \;\Psi^{\left( {\pm}\right)  }\left(  \omega\right) \; \right\rangle$, $\omega$ being the frequency.

For the $s = 1/4$ sector, the basic state is: 
\[
\left\vert \;\Psi^{\left( +\right)  }\left(  \omega\right) \; \right\rangle
\;=\;\left(  1-\left\vert \omega\right\vert ^{2}\right)  ^{1/4}\underset
{n\,=\,0,1,2..}{\sum}\frac{\left(  \omega/2\right)  ^{2n}}{\sqrt{2n\,!}}\;\left\vert
2n\right\rangle .
\]
On the other hand, the ontological states in the circle (in the limit $N\rightarrow 1$) considered by t' Hooft, and  previously proposed by London as phase states in Ref. \cite{london} in terms of the
\ eigenstates $n$ of the harmonic oscillator are:%
\begin{equation}
\left\langle \varphi\right\vert \;=\;\frac{1}{\sqrt{2\pi}}\underset{n\;=\;0,1,2..}%
{\sum}e^{i\varphi n}\left\langle n\right\vert \label{cir}%
\end{equation}
They are overcomplete:
\begin{align*}
\left\langle \varphi\right\vert \left\vert \varphi^{\prime}\right\rangle  &
\;=\;\frac{1}{2\pi}\underset{m\;=\;0,1,2...\;}{\;\sum}\underset{n\;=\;0,1,2...}{\;\sum}e^{i\varphi
n}e^{-i\varphi^{\prime}m}\left\langle n\right\vert \left\vert m\right\rangle
\\
& \;=\;\frac{1}{2\pi}\underset{n\;=\;0,1,2...}{\sum}e^{i\left(  \varphi-\varphi
^{\prime}\right)  n}%
\end{align*}
and solve the identity%
\begin{align*}
\int_{0}^{2\pi}\left\vert \varphi\right\rangle \left\langle \varphi\right\vert
d\varphi & \;=\;\underset{\delta_{\,n,m}}{\underbrace{\frac{1}{2\pi}\int_{0}^{2\pi
}e^{i\left( n-m\right)  \varphi}\,d\varphi}}\underset{\, m\;=\;0,1,2...\;}{\;\sum}%
\underset{\;n\,=\,0,1,2...\;}{\;\sum}\left\vert m\right\rangle \left\langle n\right\vert
\\
& \;= \;\overset{\infty}{\underset{n\,=\,0,1,2...}{\sum}}\left\vert n\right\rangle
\left\langle n\right\vert \;= \;\mathbb{I}%
\end{align*}

Therefore, the scalar product of the two sets of states are:
\begin{align}
\left\langle\; \varphi\right\vert \left\vert \Psi^{\left(+\right)  }\left(
\omega\right)  \;\right\rangle  &  \;=\;\frac{\left(  1-\left\vert \omega\right\vert
^{2}\right)^{1/4}}{\sqrt{2\pi}}\underset{m\,= \,0,1,2..,\,}{\sum}\underset
{n\,=\,0,1,2..}{\sum}\frac{\left(  \omega/2\right)^{2n}}{\sqrt{2n\,!}}\;e^{i\varphi
m}\left\langle m\right\vert \left\vert 2n\right\rangle \label{cso}\\
& \nonumber\\
\left\langle \;\varphi\right\vert \left\vert \Psi^{\left(+\right)  }\left(
\omega\right)\;  \right\rangle  &  =\;\frac{\left( 1-\left\vert \omega\right\vert
^{2}\right)^{1/4}}{\sqrt{2\pi}}\underset{n\,=\,0,1,2..}{\sum}\frac{\left(
\omega\, e^{i\varphi}/2\right)^{2n}}{\sqrt{2n\,!}}\nonumber\\\\
&  =\;\frac{\left(  1-\left\vert z\right\vert ^{2}\right)^{1/4}}{\sqrt{2\pi}%
}\underset{n\,=\,0,1,2..}{\sum}\frac{\left(  z/2\right)^{2n}}{\sqrt{2n\,!}},\quad\;\quad
z\,=\,\omega \, e^{i\varphi},  
\end{align}

with \,$z\,=\,\omega\, e^{i\varphi}$, the analytic function in the disc is modified by
the phase without changing the consistency of the conformal map.

\medskip

Similarly, for the sector $s=3/4$:
\begin{align}
\left\langle \;\varphi \right\vert \left\vert \Psi^{\left(  -\right)  }\left(
\omega\right) \; \right\rangle  &  \;=\;\frac{\left( 1-\left\vert \omega\right\vert
^{2}\right)^{3/4}}{\sqrt{2\pi}}\underset{n\,=\,0,1,2..}{\sum}\frac{\left(
\omega \;e^{i\varphi}/2\right)^{2n+1}}{\sqrt{\left(2n+1\right)\,!}%
}\nonumber\\\\
&  =\;\frac{\left(  1-\left\vert z\right\vert ^{2}\right) ^{3/4}}{\sqrt{2\pi}%
}\underset{n\,=\,0,1,2..}{\sum}\frac{\left(  z/2\right)^{2n+1}}{\sqrt{\left(
2n+1\right)\,!}}%
\end{align}

\bigskip

Notice that by taking the scalar product between the ontological states \,
$\left\vert \;\varphi \;\right \rangle$\, and the
states of $Mp(2)$, we obtain {\it two non-equivalent} expansions in terms of
analytical functions on the disk for the sectors of the minimal representations
$s = 1/4,\, 3/4$, corresponding to the {\it even} and {\it odd} $n$ eigenstates  of the harmonic
oscillator respectively. 

\medskip

Consequently, $\ \left( \omega\, e^{i\varphi} = z \right)$:%
\begin{subequations}
\label{0}%
\begin{equation}%
\begin{array}
[c]{c}%
\\
\left\langle \;\varphi\right\vert \left\vert \;\Psi^{(\pm)}\left(  z\right)\;
\right\rangle =\\
\end{array}
\left\{
\begin{array}
[c]{cc}%
\frac{\left(1\,-\,\left\vert z\right\vert ^{2}\right)^{1/4}}{\sqrt{2\pi}%
}\underset{n\,=\,0,1,2..}{\sum}\frac{\left(  z/2\right)^{2n}}{\sqrt{(2n)\,!}} \;\; &
\text{ (+) : even states}\\
& \\
\frac{\left(1\,-\,\left\vert z\right\vert ^{2}\right)^{3/4}}{\sqrt{2\pi}%
}\underset{n\,=\,0,1,2..}{\sum}\frac{\left(  z/2\right)^{2n+1}}{\sqrt{\left(
2n+1\right)\,!}} \;\; & \text{ (-) : odd states}%
\end{array}
\right.  \label{tabla}%
\end{equation}
\end{subequations}

\bigskip

Therefore, the total or complete projected state $\left\langle \;\varphi\right\vert \left\vert \;\Psi\left(  z\right)\;
\right\rangle$ is given by:
\begin{equation}
\left\langle \;\varphi\right\vert \left\vert \;\Psi\left(  z\right)\;
\right\rangle \;= \;\left\langle \;\varphi\right\vert \left\vert \;\Psi^{(+)}\left(  z\right)\;
\right\rangle \; +\; \left\langle \;\varphi\right\vert \left\vert \;\Psi^{(-)}\left(  z\right)\;
\right\rangle \;
\end{equation}

\begin{equation}
\left\langle \;\varphi\;\right\vert \left\vert \;\Psi\left(  z\right)\;
\right\rangle \;=\;\frac{\left(1-\left\vert z\right\vert ^{2}\right)^{1/4}%
}{\sqrt{2\pi}}\underset{n\,=\,0,1,2..}{\sum}\frac{\left( z/2\right)^{2n}}%
{\sqrt{(2n)\,!}}\left[ \, 1\;+\;\left(1-\left\vert z\right\vert ^{2}\right)
^{1/2}\frac{\left(  z/2\right)  }{\sqrt{2n+1}}\;\right]  \label{f}%
\end{equation}

\medskip

\bigskip

Now let us consider the following observations about the total expression $\left\langle \;\varphi\;\right\vert \left\vert \;\Psi\left(  z\right)\;
\right\rangle :$ 

\begin{figure}
[ptb]
\begin{center}
\includegraphics[
height=4.062000in,
width=7.831700in,
height=3.4013in,
width=6.5371in
]
{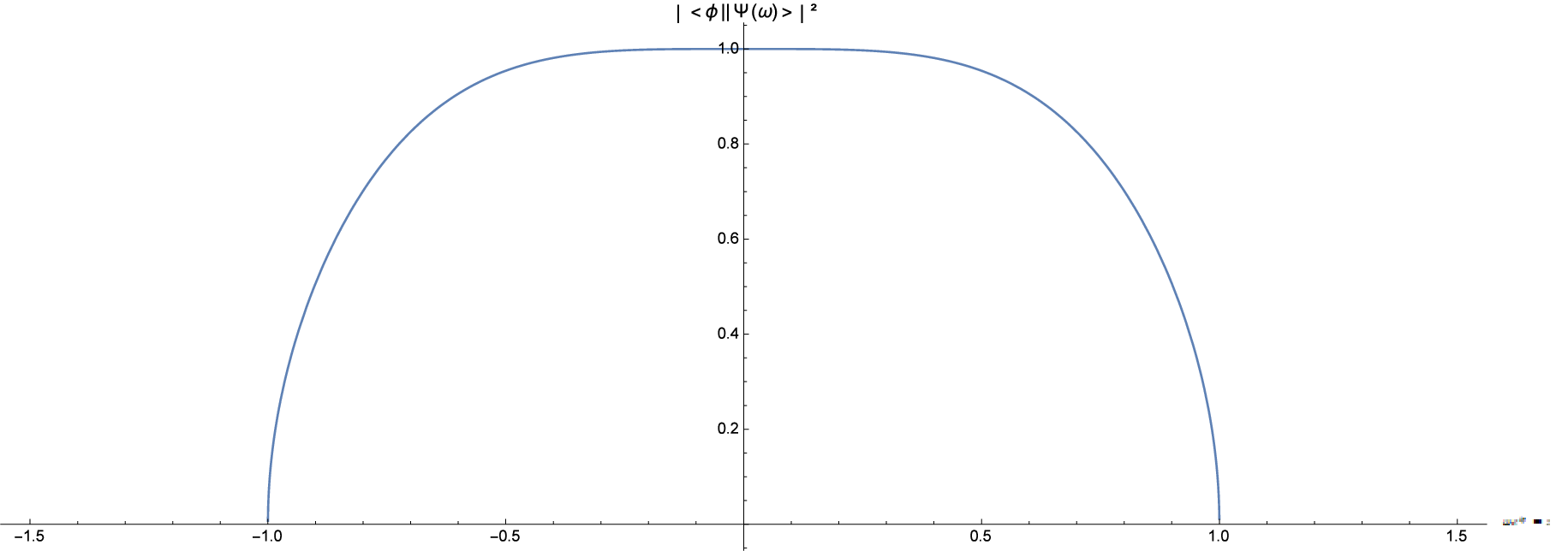}
\caption{Graphical representation of the 
the norm of the projection of the total Megaplectic $Mp(2)$ states onto the  circle $\left\vert \varphi \right\rangle
$ states: The function $\left\vert \left\langle
\varphi\right\vert \left\vert \Psi\left(  \omega\right)  \right\rangle
\right\vert ^{2}$. As shown, analyticity is evident since it clearly respects 
$\left\vert z = \omega e^{i\varphi}\right\vert = \left\vert \omega\right\vert < 1.$}
\label{f1}%
\end{center}
\end{figure}

\begin{figure}
[ptb]
\begin{center}
\includegraphics[
height=5.175000in,
width=7.991700in,
height=3.7844in,
width=5.8332in
]
{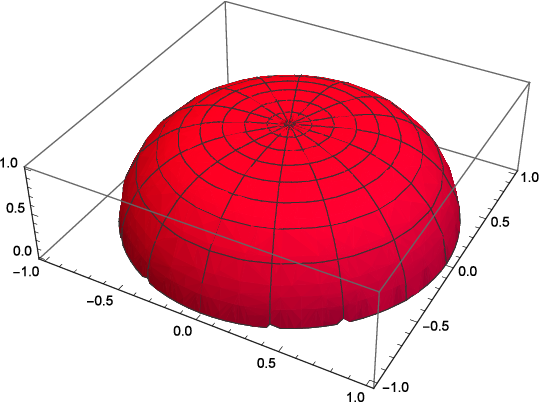}
\caption{Three-dimensional representation of the norm of the projection of the total Megaplectic $Mp(2)$ states on the  circle $\left\vert \varphi \right\rangle
$ states: The function  $\left\vert \,\left\langle\,
\varphi\right\vert \left\vert \Psi\left(  \omega\right)\,  \right\rangle \,
\right\vert^{2}$, showing clearly the analytical character due to $\left\vert
z = \omega\, e^{i\varphi} \right\vert =\left\vert \omega\right\vert < 1.$}%
\label{f2}%
\end{center}
\end{figure}

\begin{itemize}
\item{{\bf (i)} The function $\left\langle \varphi\right\vert \left\vert \Psi\left(
z\right)  \right\rangle $ is analytic on the unit disk: $\left\vert
z\right\vert\; =\;\left\vert z\right\vert < 1$.}

\bigskip

\item{{\bf (ii)} The topology of the circle induced by the state $\left\langle
\;\varphi\;\right\vert $ \,Eq. (\ref{cir}) only modifies the phase of $\omega$ (e.g:
$\omega \,e^{i\varphi} = z)$ in the projection  $\left\langle\;
\varphi \;\right\vert \left\vert \;\Psi\left(  \omega\right) \; \right\rangle$  Eq. (\ref{f}). This
is so because the topology of the circle with $R=1$ coincides with that of the
unit disk.}

\bigskip

\item{{\bf (iii)}  The norm of Eq. (\ref{f}) is obtained giving as a result the function: 

\begin{align*}
\left\vert \;\left\langle\, \varphi\,\right\vert \left\vert \Psi\left(
z\right) \; \right\rangle \;\right\vert^{2} &  \;=\; \frac{\left( 1\;-\;\left\vert
z\right\vert ^{2}\right)^{1/2}}{2\pi}\underset{n\,=\,0,1,2..}{\sum}%
\frac{\left\vert z/2\right\vert^{4n}}{(2n)!}\left[\;  1\;+\;\left(  1-\left\vert
z\right\vert^{2}\right)  \frac{\left\vert z/2\;\right\vert^{2}}{2n+1}\;\right]
\\
& \; =\;\; \frac{\left( 1\;-\;\left\vert
z\right\vert ^{2}\right)^{1/2}}{2\pi} \underset{n\,=\,0,1,2..}{\sum}\left[ \; \frac{\left\vert \, z/2 \,\right\vert ^{\,4n}}%
{(2n)!}\;+\;\left( \, 1\;-\;\left\vert \,z \,\right\vert ^{2}\right)  ^{3/2}\;\frac{\left\vert\,
z/2\,\right\vert^{\,2\,\left(  2n+1 \right)  }}{(2n+1)!}\;\right]  \\
&  \;=\;\left(  1\;-\;\left\vert \,z\,\right\vert ^{2}\,\right)  ^{1/2}\,\text{cosh}\,(\,\left\vert \, z\,\right\vert ^{2}/{2}\,)  \;+\;\left( \, 1\;-\;\left\vert \,
z \, \right\vert ^{2}\,\right)^{3/2} \, \sinh \,(\,\left\vert \,z\,\right\vert ^{2} /{2}\,)
\end{align*}

that it is evidently analytic in the disc  $\left\vert
z \right\vert =\left\vert \omega\right\vert < 1$,          graphically represented in the Figures \ref{f1} and \ref{f2}.}
\end{itemize}

\section{General Coherent States in Configuration Space}%

In this section we will elucidate and clarify the concept of the
existence of an underlying quantum structure in physical systems, in light of
the results of the previous section. For this purpose,
we implement the Minimal Group Representation approach to the case of the dynamics of a
particle in the geometry of a cylinder. The dynamics of a particle on a cilinder have been studied in Refs \cite{kowalski1}, \cite{kowalski2}, \cite{gonzalez},
and in Ref \cite {ciriloJPA} for the non-orientable case. \\

In order to introduce
the Coherent States for a quantum particle on the cylinder geometry, it is possible to
follow the Barut--Girardello construction and seek the Coherent State as the solution of
the eigenvalue equation:
\[
X\left\vert \xi \right\rangle \;=\;\xi |\left\vert \xi \right\rangle
\]

with the complex parameter $\xi$, similarly to the standard case, and
\[
X\;=\;e^{i\left(  \widehat{\varphi}\,+\,\widehat{J}\right)  }%
\]
In order to analyze the Coherent State of a particle in the cylinder in the context of the Minimal Group Representation, we express the coherent states as:

\begin{equation} \label{cilins}
\left\vert \,\xi \,\right\rangle \;=\underset{}{\underset{j\;=\;-\infty}{\overset{\infty
}{\sum}}e^{\left(  l-i\varphi\right)  j}e^{-j^{2}/2}}\,\left\vert\,
j \,\right\rangle
\end{equation}

If \;$\left\vert \,j\,\right\rangle $ $\sim$ $\left\vert \,n\,\right\rangle $\; and for the
$s\;=\;1/4$ Metaplectic states $\left\vert \,\Psi^{\left(+\right)  }\left(
\omega\right) \, \right\rangle$ : 
\;

\begin{align*}
\left\langle\, \xi \,\right\vert \left\vert \,\Psi^{\left(+\right)  }\left(
\omega\right) \, \right\rangle  & =\left(  1-\left\vert \omega\right\vert
^{2}\right)^{1/4}\overset{\infty}{\underset{m\;=\;-\infty\;}{\,\sum}}%
\underset{\,n\;=\;0,1,2...}{\;\sum}\frac{\left(  \omega/2\right)^{2n}}{\sqrt{2n\,!}%
}\;e^{\left(l-i\varphi\right)\,  m}\,e^{-2m^{2}}\left\langle m\right\vert
\left\vert 2n\right\rangle \\\\
& = \,\left(  1-\left\vert \omega\right\vert ^{2}\right)^{1/4}\underset
{n\;=\;0,1,2..}{\sum}\frac{\left(\,  \omega\, e^{\,\left( l-i\varphi\right)
}/2\right)^{2n}}{\sqrt{2n\,!}}\,e^{-2n^{2}}%
\end{align*}

Similarly, for the $s\;=\;3/4$ Metaplectic states $\left\vert \,\Psi^{\left( - \right)  }\left(
\omega\right) \, \right\rangle$ : 
\begin{align*}
\left\langle \,\xi\,\right\vert \left\vert \,\Psi^{\left(-\right)  }\left(
\omega\right) \, \right\rangle  & = \left(1-\left\vert \omega\right\vert
^{2}\right)^{3/4}\overset{\infty\;}{\underset{m\;=-\infty\;}{\sum}}%
\underset{n\;=\;0,1,2...}{\sum}\frac{\left(  \omega/2\right)^{2n+1}}%
{\sqrt{\left ( 2n+1\right) \, !}}\;e^{\left(l-i\varphi\right)\,m}\, e^{-m^{2}%
/2}\,\left\langle m\right\vert \left\vert 2n+1\right\rangle \\\\
& = \left(  1 - \left\vert \omega\right\vert ^{2}\right)  ^{3/4}\underset
{n\;=\;0,1,2...}{\sum}\frac{\left(\,  \omega \,e^{\left(l-i\varphi\right)
}/2\,\right)^{2n+1}}{\sqrt{\left( 2n+1\right)\,!}}e^{-\left(2n+1\right)
^{2}/2}%
\end{align*}

\bigskip

We see that the scalar product projections taken with the cilinder $ \left\langle\, \xi\,\right\vert  $ 
 space configuration states
are similar to the projections taken with the circle  $ \left\langle\, \varphi\,\right\vert $  phase space states, but {\it in contrast} they contain weight functions: $e^{-2n^{2}}$ and  $e^{-\left(
2n+1\right)^{2}/2}$, which {\it drastically attenuate} the scalar products when $n\rightarrow\infty $:

\begin{subequations}
\label{0}%
\begin{equation}%
\begin{array} 
[c]{c}%
\\
\left\langle \,\xi\,\right\vert \left\vert \,\Psi^{(\,\pm)}\left(  \omega\right)\,
\right\rangle = 
\end{array}
\left\{
\begin{array}
[c]{cc}
\left(  1-\left\vert \omega\right\vert ^{2}\right)  ^{1/4}\underset
{n\;=\;0,1,2...}{\sum}\frac{\left( \, \omega \,e^{\left(  l-i\varphi\right)
}/2\,\right)^{2n}}{\sqrt{2n\,!}}\,e^{\,-2n^{2}} & \text{even states}\\
& \\
\left(  1-\left\vert \omega\right\vert ^{2}\right)  ^{3/4}\underset
{n\;=\;0,1,2...}{\sum}\frac{\left( \, \omega \,e^{\,\left(  l-i\varphi\right)
}/2\,\right)^{2n+1}}{\sqrt{\left(  2n+1\right)  \,!}}\;e^{-\,\left(  2n+1\right)
^{2}/2} & \text{odd states}%
\end{array}
\right. 
\end{equation}

\bigskip

Consequently, for the total state :
$$ \left\vert \,\Psi\left(  \omega\right) \,\right\rangle \;=\; \left\vert \,\Psi^{(+)} \left(  \omega\right) \,\right\rangle \; + \; \left\vert \, \Psi^{\,(-)} \left(  \omega\right)\,\right\rangle, $$ 
we have:%
\end{subequations}
\begin{equation}   \label{wfcs}
\left\langle \,\xi\,\right\vert \left\vert \,\Psi\left(  \omega\right)
\,\right\rangle =\left(  1-\left\vert \omega\right\vert ^{2}\right)
^{1/4}\underset{n\;=\;0,1,2...}{\sum}\frac{\left(  \omega\, e^{\left(  l-i\varphi
\right) }/2\right)^{2n}}{\sqrt{2n\,!}}\,e^{-2\,n^{2}}\left[  1+\left(
1-\left\vert \omega\right\vert ^{2}\right)^{1/2}\,\frac{\omega\, e^{\,\left(
l-i\varphi\right)  }}{\sqrt{2n+1}}\;e^{-(2n+1)/2}\,\right]
\end{equation}
 
 \bigskip

We discuss the implications of these results in the next Section.

\section{Implications of the Minimal Group Representation Reduction}

As we have seen so far, in order to interpret the 
dynamical scenario connected with an inherent quantum structure, the use of the London circle states takes a
true dimension only when the system is subjected to the 
minimal group representation under the action of the
metaplectic group $Mp(n)$. Let us recall that $Mp(n)$ covers $Sp(n)$ twice and in
certain cases its Hermitian structure can be extended to $OSp(n)$.

\bigskip

Below we outline some implications that do result from the developments and
analysis in the previous Sections.

\medskip

\begin{itemize}
\item {{\bf(i)} The London states (ontological states in t'Hooft's description) { \it  completely classicalize}  the inherent quantum structure {\it only} under the application of the Minimal Group Representation with  the $Mp(n)$ group taking the main role.}

\item{{\bf(ii)} The action of the Metaplectic group on the "ontological" (London) states
breaks the invariance under time reversal assumed for the dynamics of the
particle in the circle (arrow of time).}

\item{{\bf(iii)} In the case of the coherent states of a particle in the cylinder  (configuration space) of Section V, we can set in our analysis the parameter
$l = 0$ : Thus we can also assign to them the variable $ 
z \,=\, \omega\, e^{-i\varphi}$ as in the case of the particle states in a circle, (London states, phase space).}

\item{{\bf(iv)} In the case of the cilinder states, item (iii) with $l = 0$, the norm of the projection 
Eq. (\ref{wfcs}) is easily
calculated giving as a result the function: %
\begin{gather*}
\left\vert \left\langle \,\xi\,\right\vert \left\vert \Psi\left(  \omega\right)
\,\right\rangle \right\vert ^{2}\;= \underset{n\;=\;0,1,2...}{\sum}\frac{\left\vert \,
\omega/2\,\right\vert ^{4n}}{2n\,!}\;e^{-2n^{2}}\,G\left(  \omega,\varphi\right) 
\\
\\
G\left(  \omega,\varphi\right)  \;=\;\left[  1\;+\;\frac{\left(  1-\left\vert
\omega\right\vert ^{2}\right)  ^{1/2}}{\sqrt{2n+1}}\,e^{-2n-1/2} \left(
2\operatorname{Re}\left(  \omega \,e^{-i\varphi}\right)+ e^{-2n-1/2}\left(
1-\left\vert \omega\right\vert ^{2}\right)  ^{1/2}\frac{\left\vert
\omega/2\right\vert ^{2}}{\sqrt{2n+1}}\right)  \right]
\end{gather*}
\\
where we see the very fast decreasing of the function due to the exponentials $e^{-2n^{2}}$,  $e^{-(2n+1/2)}$ and $e^{-(2n+1/2)^{2}}$, arising in the projections of the Metaplectic states $\left\vert \Psi\left(  \omega\right)
\,\right\rangle$ on the cilinder states $ \left\vert \,\xi(\varphi)\,\right\rangle $ in {\it configuration space}.} 

\end{itemize}

\subsection{ The Generalized Wigner function}

\begin{figure}
[ptb]
\begin{center}
\includegraphics[
height=5.508900in,
width=8.194100in,
height=3.2517in,
width=4.8291in
]
{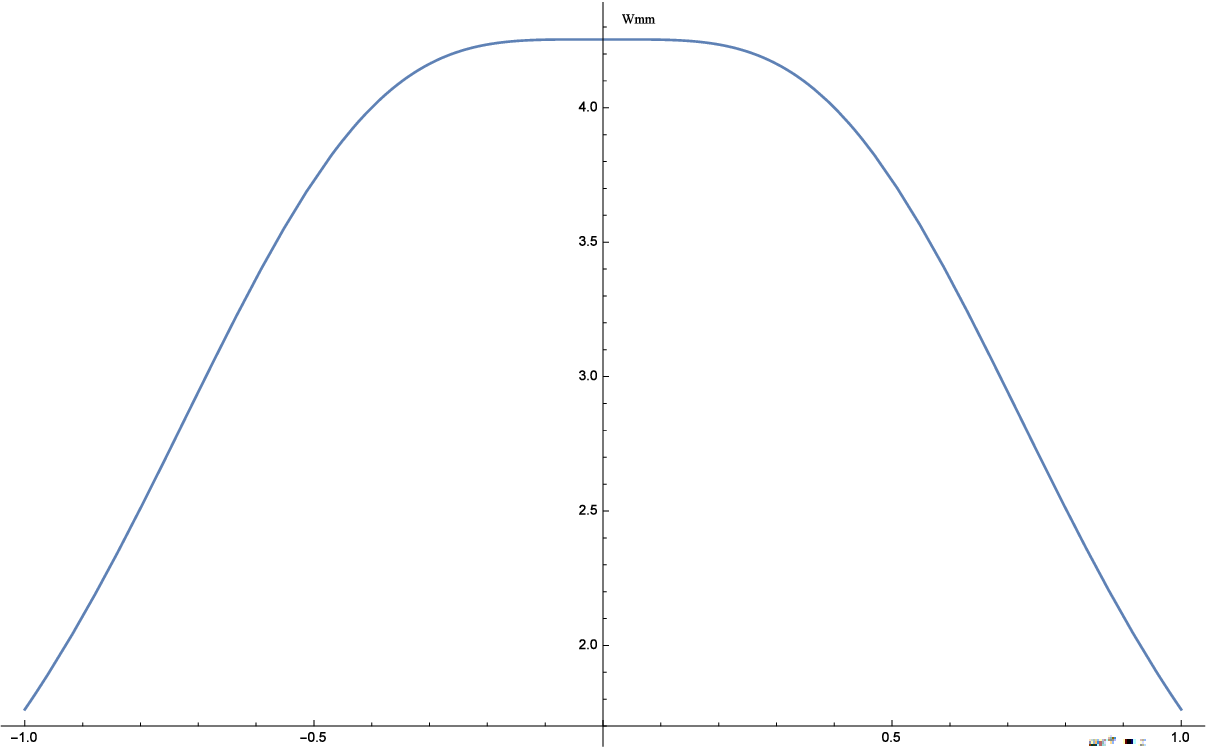}
\caption{Graphical representation of the generalized Wigner function for the
approximate $W_{mm}$ case: the shape of this distribution appears more bell-shaped
than the function in Figure 1 (square norm).}%
\label{f3}%
\end{center}
\end{figure}

Let us recall the Wigner function definition %
\[
W\left(q,p\right) \; =\;\int dv\;e^{(-pv/\hbar)}\;\,\Psi_{\omega}^{\ast}\,\left(
q-\frac{v}{2}\right) \; \Psi_{\omega}\left(  q+\frac{v}{2}\right)
\]
where $\left( q,\,p\right)$ are the  position and momenta as usual, or in general any canonical conjugate pair of variables.  In our case $(m = 1 = r = \hbar),$ the position and momentum are $\left(
\varphi,j\right)$ :
$$ \left(  q,\,p\right)\:              \rightarrow \;\left(
\varphi,j\right),$$ 
and expressing $ W $ in function of the complex
variable $z$ as before:
\[
 z = \omega \,e^{i\varphi},
\]
we obtain the following generalized Wigner function:%

\[
W_{mn}\left(z,z^{\ast}\right)  = \frac{1}{2\pi}\int d^{2}\eta\;\mathcal{M}\,(\, z_+\,)\text{ }\overset{\infty}{\underset{
m\,,\,n\,= \,0,1,2...}{\sum}} \frac{\left(\,  z_{\,+\,}\,/2\,\right)^{2n}}{\sqrt{
2n\,!}} \frac{\left(z^{\ast}_{\,-\,}\,/{2}\,\right)^{2m}}{\sqrt{
2m\,!}} \mathcal{F} \left(\,z_+\,\right)  \mathcal{F} \left(\,z^{\ast}_{\,-\,}\right)
\]
where
\[
z_{{\,\pm\,}}\; \equiv \; z\;\pm\; \eta\,/\,2
\]
\[
\mathcal{M}{\,\left( \,z\,+\,\eta/2\,\right) }\;\mathcal{\;=\;}\left( \, 1 -\left\vert\,
z + \eta\,/\,2\;\right\vert^{2}\,\right)  ^{1/2}\;\exp{[\,-(z-z^{\ast})(\eta+\eta^{\ast})/2\,]}%
\]
and
\begin{align*}
\mathcal{F} \left( z + \eta/2\right)   &  \;=\;\left[ \, 1 \,+ \,\left(\, 1-\left\vert\;
z \,+ \,\eta/2\;\right\vert^{2}\;\right)  ^{1/2}\;\frac{\left(\, z \,+\, \eta/2\,\right)  }%
{2\sqrt{2n+1}}\,\right]  \\
& \\
\mathcal{F}\left(z^{\ast}\,-\,\eta^{\ast}/2\right)   & \; =\;\left[ \; 1 + \,\left(\,
1 - \,\left\vert \;z \,+ \,\eta/2\;\right\vert ^{2}\;\right)^{1/2}\;\frac{\left(\,z^{\ast
}\,-\,\eta^{\ast}/2\,\right)  }{2\sqrt{2n+1}}\;\right]
\end{align*}\\
Notice that the complex variable is introduced in order to see the analytical
conditions of the function in a little more detail.\\
Despite the degree of
complexity of the function, we can provide an approximation in the case of $m = n$
(up to the leading terms inside the unitary disc e.g. $\left\vert\,
\eta \,\right\vert  < 1 \,)$: 

\begin{equation}
W_{m\,m}\left(  z,z^{\ast}\right)  \;\approx \; 2\,e^{-4\,\left\vert z\right\vert ^{2}%
}\left(  2\,\operatorname{Ei}\,(4 \,\left\vert \,z \,\right\vert ^{2}\,) \;+ \;\ln\frac
{1}{\left\vert\, z \,\right\vert ^{4}}+\cdot\cdot\cdot\right)  \label{wmm}
\end{equation} \\
represented in Figure \ref{f3}, Ei being the  exponential integral function. This Wigner function  for the circle states displays a classicalized typical shaped gaussian distribution, more bell- shaped that the square norm function of the states displayed in Figure 1. 

\bigskip

\section{Coset Coherent states for the circle}

\subsection{Coset coherent states}

Let us remind the definition of coset coherent states starting from a vector
$\phi_{0}$ invariant under the stability subgroup namely
\begin{equation}
H_{0}=\left\{  g\in G\mid\mathcal{U}\left(  g\right)  \phi_{0}=\phi
_{0}\right\}  \subset G. \label{tag:1}%
\end{equation}
We can see that the orbit of $\phi_{0}$ is isomorphic to the coset, e.g.
\begin{equation}
\mathcal{O}\left(  \phi_{0}\right)  \simeq G/H_{0}. \label{tag:2}%
\end{equation}
On the other hand, if we remit to the operators, e.g.
\begin{equation}
\left\vert \phi_{0}\right\rangle \left\langle \phi_{0}\right\vert \;
\equiv\;\widehat{\rho}_{0} \label{tag:3}%
\end{equation}
then the orbit now is represented as
\begin{equation}
\mathcal{O}\left(  \widehat{\rho}_{0}\right) \; \simeq \; G/H \label{tag:4}%
\end{equation}
with
\begin{align}
H  &  \;= \;\left\{g\in G\mid\mathcal{U}\left(  g\right)  \phi_{0}=\theta\phi
_{0}\right\} \nonumber\label{tag:5}\\
&  \;= \;\left\{  g\in G\mid\mathcal{U}\left(  g\right)  \widehat{\rho}%
_{0}\;\mathcal{U}^{\dagger}\left(  g\right)  \;=\;\widehat{\rho}_{0}\right\}
\subset G
\end{align}
The orbits are identified with cosets spaces of $G$ with respect to the
corresponding stability subgroups $H_{0}$ and $H$, that in the second case is
defined within a phase.\\

{\bf Quantum viewpoint}: From the quantum viewpoint: $\left\vert \phi_{0}\right\rangle \in\mathcal{H}$ (the Hilbert space) and $\rho_{0}%
\in\mathcal{F}$ (the Fock space) are $V_{0}$ normalized fiducial vectors
(embedded unit sphere in $\mathcal{H}$, and its real dimension is less than or
equal to the dimension of $G$.). \\

{\bf Coherent state:} A generalized coherent state system is now
defined as the collection of unit vectors $\phi(g)$ comprising the orbit
$\mathcal{O}\left(  \phi_{0}\right)  $; thus, it brings together $[\;\mathcal{H}$%
, $G$, $\mathcal{U}\left( g\right)  $ and $\phi_{0}\;]$ in a special way, namely:
\[
\mathcal{U}\left(  G/H_{0}\right)  \phi_{0}\;=\;\phi\left(  g\right)
\]

\subsection{Geometry of the group and coset}

As a first step in the construction of the coset coherent states on the circle
we take an element of the Euclidean group $E\left(  2\right)$ in a matrix representation derived
from the exponentiation of the generators of the algebra times the
representative parameters of the coordinates, namely:%
\[
E\left(2 \right) = \left\{\left(\;\;
\begin{array}
{ccccc}%
\cos\varphi &  & -\sin\varphi &  & x\\
&  &   \\
\sin\varphi &  & \cos\varphi &  & y\\
&  &   \\
0 &  & 0 &  & 1
\end{array}
\right)  \mid x,y\in\mathbb{R}^{2},\;\varphi\in S^{1} \;\right\}
\]

\bigskip

As can be seen, the circle (geometrically and topologically) is perfectly
described by the coset $G/H = E\left(  2\right)  /\mathbb{T}_{2}$ with
$\mathbb{T}_{2}$ being the group of translations in the plane as a stability subgroup.

\subsection{Maurer-Cartan forms and vector fields}

Let us to consider 
$$E\left(  2\right)  ^{-1}=\left(
\begin{array}
[c]{ccccc}%
\cos\varphi &  & \sin\varphi &  & -x\cos\varphi-y\sin\varphi\\
&  &  &  & \\
-\sin\varphi &  & \cos\varphi &  & x\sin\theta-y\cos\varphi\\
&  &  &  & \\
0 &  & 0 &  & 1
\end{array}
\right)$$

Then we can obtain the Maurer-Cartan forms via pullback e.g.%
\[
E\left(  2\right)  ^{-1}\;dE\left(  2\right)  \;=\;\omega^{\varphi}g_{\varphi
}\;+\;\omega^{x}g_{x}\;+\;\omega^{y}g_{y}%
\]
where \;$(g_{\varphi},\,g_{x},\,g_{y})$ \, are the generators of the \ respective algebra,
namely:

\[
g_{\varphi}=\left(
\begin{array}
[c]{ccc}%
0 & -1 & 0\\
1 & 0 & 0\\
0 & 0 & 0
\end{array}
\right)  ,\text{ \ \ \ \ \ }g_{x}=\left(
\begin{array}
[c]{ccc}%
0 & 0 & 1\\
0 & 0 & 0\\
0 & 0 & 0
\end{array}
\right)  ,\text{ \ \ \ }g_{y}=\left(
\begin{array}
[c]{ccc}%
0 & 0 & 0\\
0 & 0 & 1\\
0 & 0 & 0
\end{array}
\right)
\]
Then, we have
\begin{align*}
\omega^{\varphi} &  \;=\;d\varphi\\
\omega^{x} & \; =\;\cos\varphi \;dx\;+\;\sin\varphi \;dy\\
\omega^{y} &  \;=\;-\sin\varphi \;dx\;+\;\cos\varphi \;dy
\end{align*}
with the Cartan structure equations showing geometrically the closing of the
E$\left(  2\right)  $ algebra, namely,
\begin{align*}
d\omega^{\varphi} & \; = \;0 \;= \;\omega^{x}\wedge\omega^{y}\;=\;dx\wedge dy\\
d\omega^{x} &  \;=\;\omega^{\varphi}\wedge\omega^{y}\\
d\omega^{y} &  \;=\;\omega^{x}\wedge\omega^{\varphi}%
\end{align*}

From the above results, the vector fields can be computed in the standard
manner
\begin{align*}
e_{\varphi}  & \; =\;\partial_{\varphi}\\
e_{x}  &  \;=\;\cos\varphi\;\partial_{x}\;+\;\sin\varphi\;\partial_{y}\\
e_{y}  & \; =\;-\sin\varphi\;\partial_{x}\;+\;\cos\varphi\;\partial_{y}%
\end{align*}
with the commutation relations%
\[
\left[  e_{\varphi},\, e_{x}\right]  =\;e_{y},\text{ \ \ \ \ \ \ \ \ \ \ }\left[
e_{\varphi},\, e_{y}\right] \; =\;-\,e_{x},\text{ \ \ \ \ \ \ \ \ \ \ }\left[
e_{x},\, e_{y}\right] \; =\;0
\]

\subsection{Coset coherent states}

The steps to follow for the determination of the coset coherent states are the following:\\

 (i) The Coset G/H identification \;$\rightarrow E\left(  2\right)  /\mathbb{T}_{2}$,
\,  being $\mathbb{T}_{2}$ the group of translations $\{g_{x},g_{y}%
\}\in\mathbb{T}_{2}$.

(ii) The Fiducial vector determination: It is annihilated by all the generators $h$
of the stability subgroup $H$ and for instance, invariant under the action of
$H.$  We propose:%
\[
\left\vert \;A_{0}\;\right\rangle \;=\;A\left(  \varphi,x,y\right) \; \left\vert
\varphi\right\rangle
\]
where $\left\vert \varphi\right\rangle $ is the London (circle) state, that is
expanded in the $\left\vert n\right\rangle $ state of the harmonic oscillator and
\[
A\left(  \varphi,x,y\right)_{(\pm)}\;  =\;\left(  \cos\varphi\;\pm\;\sin\varphi\right)
x\;+\;\left(  \mp\cos\varphi\;+\;\sin\varphi\right)y,
\]
such that we can see:
\[
\left(  e_{x}\;+\; e_{y}\right)  A\left(  \varphi,x,y\right)_{(\,\pm)} \;= \; 0
\]

\medskip

(iii) The coherent state is defined as the action of an element of the coset group on
the fiducial vector $\left\vert A_{0}\right\rangle$, consequently the coherent
state (still unnormalized yet) takes the form:\\
\begin{align}
e^{-\alpha\,\partial_{\varphi}}\left\vert A_{0}\right\rangle  &  =\frac{1}{\sqrt
{2\pi}}\left(  A_ + \cos \alpha + A_{-} \sin \alpha\right)  \underset{n\,=\,0,1,2..}{\sum
}e^{-\alpha\,\partial_{\varphi}}e^{-i\,\varphi\, n}\left\vert \,n\,\right\rangle
\label{CCS}\\
\left\vert \;\alpha,\varphi\right\rangle  &  =\frac{1}{\sqrt{2\pi}}%
\mathcal{S}\left(  \alpha,\varphi\right)  \underset{\left\vert \varphi
-\alpha/2\right\rangle }{\underbrace{\underset{n = 0,1,2..}{\sum}e^{-i\left(
\varphi-\alpha/2\right)  n}\left\vert n\right\rangle }}\nonumber\\
&  = \frac{1}{\sqrt{2\pi}}\;\mathcal{S}\left(  \alpha,\varphi\right)  \;\left\vert
\varphi-\alpha/2\;\right\rangle
\end{align}\\
where $\alpha\in
\mathbb{C}$ is an arbitrary complex parameter in the element of the coset, 
 which must be adjusted after normalization, being%
\[
\mathcal{S}\left(  \alpha,\varphi\right)\;  \equiv \;\left(  A_{+}\cos\alpha+A_{-}%
\sin\alpha\right)
\]

Some observations with respect to the
expression Eq.(\ref{CCS}) that is the most general coherent state expression from the Klauder-Perelomov
construction viewpoint are the following: \\
Notice that the state can be normalized from the
overlap taking the form:
\begin{align*}
\left\langle \beta,\varphi^{\prime}\right\vert \left\vert \alpha
,\varphi\right\rangle  &  =\frac{1}{2\pi}\mathcal{S}\left(  \beta^{\ast
},\varphi^{\prime}\right)  \mathcal{S}\left(  \alpha,\varphi\right)
\underset{m = 0,1,2..\,}{\sum}\underset{\,n = 0,1,2..}{\sum}e^{i\,\left(  \varphi
^{\prime}-\beta^{\ast}/2\,\right)\,  m}e^{-i\left(  \varphi-a/2\,\right)  \,n}\left\langle
m\right\vert \left\vert n\right\rangle \\
&  = \frac{1}{2\pi}\mathcal{S}\left(  \beta^{\ast},\varphi^{\prime}\right)
\mathcal{S}\left(  \alpha,\varphi\right)  \underset{n \,= \, 0, 1, 2..}{\sum
}e^{- i\,\left[  \varphi-\varphi^{\prime}-\left(  \alpha-\beta^{\ast}\right)
/2\right]\,  n}\\
&  =\;\frac{1}{2\pi}\;\;\frac{\mathcal{S}\left(  \beta^{\ast},\varphi^{\prime
}\right) \; \mathcal{S}\left(  \alpha,\varphi\right)  }{1-e^{-\,i\,\left(
\,\varphi-\varphi^{\prime}\,-\,\left(  \alpha\,-\,\beta^{\ast}\right)  /2\,\right)}}%
\end{align*}
Then,  the state is fully normalizable: $\varphi\rightarrow\varphi^{\prime}$ 
Iff the parameter $\alpha$ have $\operatorname{Im}\alpha \;\neq \;0$%

\begin{equation}
\left\vert \,\left\vert \alpha,\varphi\right\rangle \,\right\vert^{2}\;= \;\frac
{1}{2\pi}\;\;\frac{\mathcal{S}\left(\alpha^{\ast},\varphi\right)\;  \mathcal{S}%
\left(  \alpha,\varphi\right)  }{1-e^{-\,i\left( \, \alpha^{\ast}\,-\,\alpha\right)
/2}} \label{qn}%
\end{equation}
where:
\[
\mathcal{S}\left(  \alpha^{\ast},\varphi\right)  \mathcal{S}\left(
\alpha,\varphi\right) =\left(  x^{2} + y^{2}\right)  \cosh\left(
2\operatorname{Im}\alpha\right)  -\left(  x^{2}\,-\,y^{2}\right)  \sin2\left(
\operatorname{Re}\alpha-\varphi\right)  \,+\,2xy\cos2\left(  \operatorname{Re}%
\alpha-\varphi\right)
\]
Consequently, it solves the problem of the London states that are overcomplete but
clearly not normalizable when $\varphi\rightarrow\varphi^{\prime}$:
\begin{equation}   \label{qn2}        \left\langle \varphi\right\vert \left\vert \varphi^{\prime}\right\rangle
\;=\;\frac{1}{2\pi}\underset{n\,=\,0,1,2..}{\sum}e^{i\left(  \varphi-\varphi^{\prime
}\right)  n}\;=\;\frac{1}{2\pi}\frac{1}{1-e^{-i\left(  \varphi-\varphi^{\prime
}\right)  }}          %
\end{equation}

We see explicitely from these expressions Eq.(\ref{qn}), Eq.(\ref{qn2})  how the general coherent states on the circle $\left\vert \alpha,\varphi\right\rangle$ (with the coherent characteristic complex parameter $\alpha$) solve the problem of the non normalizability of the known (London, 't Hooft) $\left\vert \,\varphi\right\rangle$  states in the circle. 

From Eq.(\ref{qn}) the normalized state coherent state $\left\langle \varphi\right\vert \left\vert \varphi^{\prime}\right\rangle $  is:

\begin{equation}
\left\vert \alpha,\varphi\right\rangle \;=\; \underset{\mathcal{N}}{\underbrace
{\sqrt{1-e^{-\operatorname{Im}\alpha}}\;e^{i\arg\mathcal{S}}}}\underset
{n\,=\,0,1,2..}{\sum}e^{-i\left(  \varphi-\alpha/2\right) \, n}\;\left\vert
n\right\rangle \label{ccs}%
\end{equation}

\medskip

However, the identity is not resolved in a strict sense, but in a weak sense,
always for $\operatorname{Im}\alpha>0$\,:

\begin{align*}
\int_{0}^{2\pi}\left\vert \alpha,\varphi\right\rangle \left\langle
\alpha,\varphi\right\vert \, d\varphi &  =\frac{1}{2\pi}\int_{0}^{2\pi}%
\underset{n\,=\,0,1,2..\,}{\,\sum}e^{i(m-n)\varphi}e^{i\left(  \alpha n-\alpha^{\ast
}m\right)  /2}\;d\varphi\underset{m\,= \,0, 1, 2...\,}{\sum}\underset{n \,=\, 0, 1, 2..\,}{\,\sum
}\left\vert m\right\rangle \left\langle n\right\vert \\
&  = \overset{\infty}{\underset{n \,=\, 0, 1, 2...\,}{\,\sum}}e^{-n\operatorname{Im}\alpha
}\left\vert n\right\rangle \left\langle n\right\vert = \left(
\begin{array}
[c]{ccccc}%
1 &  &  &  & \\
& e^{-\operatorname{Im}\alpha} &  &  & \\
&  & e^{-2\operatorname{Im}\alpha} &  & \\
&  &  & \ddots & \\
&  &  &  & e^{-n\operatorname{Im}\alpha}%
\end{array}
\right)
\end{align*}

and which clearly shows the role 
played by the complex characteristic coherent state parameter $\alpha$.

\section{ Action of the Mp(2) Group on the coset coherent states on the circle}

Again, et's look at the sector $s = 1/4$ of the Hilbert space spanned by the $Mp(2)$
coherent states (unnormalized) , the basic state is%
\[
\left\vert \Psi^{\left(  +\right)  }\left(  \omega\right)  \right\rangle
=\left(  1-\left\vert \omega\right\vert ^{2}\right)  ^{1/4}\underset
{n=0,1,2..}{\sum}\frac{\left(  \omega/2\right)  ^{2n}}{\sqrt{2n!}}\,\left\vert\,
2n\, \right\rangle
\]

On the other hand:%
\begin{equation}
\left\langle \alpha,\varphi\right\vert =\mathcal{N}^{\ast}\underset
{n=0,1,2..}{\sum}e^{i\left(  \varphi-\alpha^{\ast}/2\right)  n}\,\left\langle
\,n \,\right\vert
\end{equation}
Then, with all the definitions above, an in principle excluding the
normalization $\mathcal{N}$ we have%
\begin{align}
\left\langle \alpha,\varphi\right\vert \left\vert \Psi^{\left(  +\right)
}\left(  \omega\right)  \right\rangle  &  =\frac{\left(  1-\left\vert
\omega\right\vert ^{2}\right)  ^{1/4}}{\sqrt{2\pi}}\underset{m=0,1,2..}{\sum
}\underset{n=0,1,2..}{\sum}\frac{\left(  \omega/2\right)  ^{2n}}{\sqrt{2n!}%
}e^{i\left(  \varphi-\alpha^{\ast}/2\right)  m}\left\langle m\right\vert
\left\vert 2n\right\rangle \\
& \nonumber\\
\left\langle \alpha,\varphi\right\vert \left\vert \Psi^{\left(  +\right)
}\left(  \omega\right)  \right\rangle  &  =\frac{\left(  1-\left\vert
\omega\right\vert ^{2}\right)  ^{1/4}}{\sqrt{2\pi}}\underset{n=0,1,2..}{\sum
}\frac{\left(  \omega e^{i\left(  \varphi-\alpha^{\ast}/2\right)  }/2\right)
^{2n}}{\sqrt{2n!}}=\nonumber\\
&  =\frac{\left(  1-\left\vert \omega\right\vert ^{2}\right)  ^{1/4}}%
{\sqrt{2\pi}}\underset{n=0,1,2..}{\sum}\frac{\left(  z^{\prime}/2\right)
^{2n}}{\sqrt{2n!}}%
\end{align}
with
\[
\omega \,e^{\,i\left(  \varphi\,-\,\alpha^{\ast}/2\,\right)  }\;=\;z\; e^{\,-i\,\alpha^{\ast}%
/2\,}\;\equiv\; z^{\prime}%
\]
$:$ the analytic function in the disc is now modified by the complex phase
$(\,\varphi-\alpha^{\ast}/2\,)$.\\

Similarly, for the sector $s = 3/4$ of the $Mp(2)$ states we have:

\begin{align}
\left\langle \alpha,\varphi\right\vert \left\vert \Psi^{\left(  -\right)
}\left(  \omega\right)  \right\rangle  &  =\frac{\left(  1-\left\vert
\omega\right\vert ^{2}\right) ^{3/4}}{\sqrt{2\pi}}\underset{n=0,1,2..}{\sum
}\frac{\left(  \omega e^{i\left(  \varphi-\alpha^{\ast}/2\right)  }/2\right)
^{2n+1}}{\sqrt{\left(  2n+1\right)  !}}\nonumber\\
&  =\frac{\left(  1-\left\vert \omega\right\vert ^{2}\right)  ^{3/4}}%
{\sqrt{2\pi}}\underset{n=0,1,2..}{\sum}\frac{\left(  z^{\prime}/2\right)
^{2n+1}}{\sqrt{\left(  2n+1\right)  !}}%
\end{align}

\bigskip

Notice that by taking the scalar product between the coset coherent
state  $\left\langle \alpha,\varphi\right\vert$  and the coherent states of $Mp(2)$, $\left\vert \Psi^{\left(  -\right)
}\left(  \omega\right)  \right\rangle $ we obtain two non-equivalent expansions in
terms of analytical functions on the disk for the sectors of the minimal
representations $s = 1/4 ,3/4$, {\it even and odd $n$ states}, respectively, in the eigenstaes $ \left\vert \,n \,\right\rangle $ of the
harmonic oscillator. \\
Consequently $\ \left(\;  \omega \,e^{\,i\, \left(\, 
\varphi\alpha^{\ast}/2\, \right)}\;\equiv \;z^{\prime}\;\right)  $:
\begin{subequations}
\label{0}%
\begin{equation}%
\begin{array}
[c]{c}%
\\
\left\langle \,\alpha,\varphi\right\vert \left\vert \Psi ^{(\pm)}\left(  z^\prime\right)\,
\right\rangle = \\
\end{array}
\left\{
\begin{array}
[c]{cc}%
\left(  1-\left\vert z^\prime\right\vert ^{2}\right)  ^{1/4}\underset
{n=0,1,2..}{\sum}\frac{\left(  z^{\prime}/2\right)  ^{2n}}{\sqrt{2n!}} &
\text{\;\; (+): \;even states}\\
& \\
\left(  1-\left\vert z^\prime \right\vert ^{2}\right)  ^{3/4}\underset
{n=0,1,2..}{\sum}\frac{\left(  z^{\prime}/2\right)  ^{2n+1}}{\sqrt{\left(
2n+1\right)  !}} & \text{\;\;(-):\;\;odd states}%
\end{array}
\right.
\end{equation}\\
Therefore, for the total projected state,   $\left\langle \; \alpha , \varphi\, \right\vert \left\vert \;\Psi\left(  z^\prime \right)\;
\right\rangle$ :

\begin{equation}
\left\langle \; \alpha , \varphi\, \right\vert \left\vert \;\Psi\left(  z^\prime \right)\;
\right\rangle= \;\left\langle \; \alpha , \varphi\, \right\vert \left\vert \;\Psi^{(+)}\left(  z^\prime \right)\;
\right\rangle \; +\; \left\langle \; \alpha , \varphi\, \right\vert \left\vert \;\Psi^{(-)}\left(  z^\prime \right)\;
\right\rangle \;,
\end{equation}

We have%

\end{subequations}
\begin{equation}
\left\langle \,\alpha,\varphi\right\vert \left\vert \Psi\left(  z^\prime \right)\,
\right\rangle =\left(  1-\left\vert z^\prime \right\vert ^{2}\right)
^{1/4}\underset{n=0,1,2..}{\sum}\frac{\left(  z^{\prime}/2\right)  ^{2n}%
}{\sqrt{2n!}}\left[  1+\left(  1-\left\vert z^\prime\right\vert ^{2}\right)
^{1/2}\frac{\left(  z^{\prime}/2\right)  }{\sqrt{2n+1}}\right]  \label{gg}%
\end{equation}
\\
We now consider the following observations:\\

{\bf(i)} The analiticity condition of the function $\left\langle \; \alpha , \varphi\, \right\vert \left\vert \Psi\left(  z^\prime \right)  \right\rangle $ on the disk now
constrained, taking into account: $\left\vert \,z^{\prime}\right\vert \,= \,\left\vert z \right\vert e^{-\operatorname{Im}\alpha/2}\,<\,1 $ which occurs under the already accepted condition of arising from
the normalization function.\\

{\bf (ii)} The topology of the circle induced by the coset coherent state
$\left\langle \alpha,\varphi\right\vert $ Eq. (\ref{ccs}) not only modifies the
phase of $\omega$ (e.g: $\omega \,e^{i\,\left( \, \varphi-\alpha^{\ast}/2 \,\right)
} = z^{\prime}\,)$ but also the ratio of the disc due the displacement generated
by the action of the coset.\\

{\bf(iii)} The norm square of Eq. 
  (\ref{gg}) is easily calculated giving as a result
the function:

\begin{align*}
\left\vert \;\left\langle \varphi \right\vert \left\vert \Psi\left(
z^\prime \right)  \right\rangle \;\right\vert ^{2} &  \; = \;\left(  1 -\left\vert
\,z^{\prime}\right\vert^{2}\right)^{1/2} \;  \cosh\left(  \frac{\left\vert z^{\prime}\right\vert ^{2}
}{2}\right)\; + \;\left(  1-\left\vert z^\prime \right\vert
^{2}\right)^{3/2} \, \sinh \left(  \frac{\left\vert z^{\prime}\right\vert ^{2}}{2} \right)\;  + \\
& + \;  \left(  1 -\left\vert
\, z^{\prime}\right\vert^{2}\right)^{1/2} \;  \operatorname{Re}\,( z^{\prime}) \underset{n\;=\;0, 1 , 2..}{\sum}\frac{\left\vert
z^{\prime}/2\right\vert ^{ \,4n}}{2n\,! \,\left(  2n+1\right)} , \;\;\; \text{\,$ 
z^{\prime} \,=\,\omega \,e^{\,i\,\left(\,  \varphi -\alpha^{\ast}/2\,\right)}$}
\end{align*}\\
with a decreasing tail as $n$ increases, and 
showing the analicity, in this case in the disc $ \left\vert z^\prime \right\vert\; < \;1 $, 
with the same comments as in the items (i)-(ii) above.

\section{Concluding remarks}

In this paper relevant implications inherent to the description of quantum
theory, in particular when it takes a classical aspect, were elucidated and discussed from the
principle of minimum group representation. To
this end the concept of classical quantum duality was considered demonstrating
the non-existence of ontological or hidden variables in the reality of the
physical scenario analyzed.\\

{\bf (1)} The application of the Minimal Representation Group Principle (MGRP) to the London state (circle-phase states), e.g. the
application of an element of the $Mp(2)$ group on the London state, {\it immediately
 classicalizes} the physical scenario considered: the quantum dynamics in the circle takes on the classical character. \\

{\bf (2)} The application of the MGRP to the London states {\it naturally} introduces the
analytic functions through the action of the basic  (coherent,
$s = 1/4,\, 3/4$) states of the Metaplectic group in the Bargmann representation.  This is in 
contrast with the case of t' Hooft in Ref \cite{tHooft1}  where although similar functions and results in the circle are considered,  the analyticity in the unit disk $ \left\vert\, z \,\right\vert \,<\, 1 $, is introduced differently.\\

\begin{figure}
[ptb]
\begin{center}
\includegraphics[
height=3.399000in,
width=6.920200in,
height=4.4022in,
width=7.4648in
]
{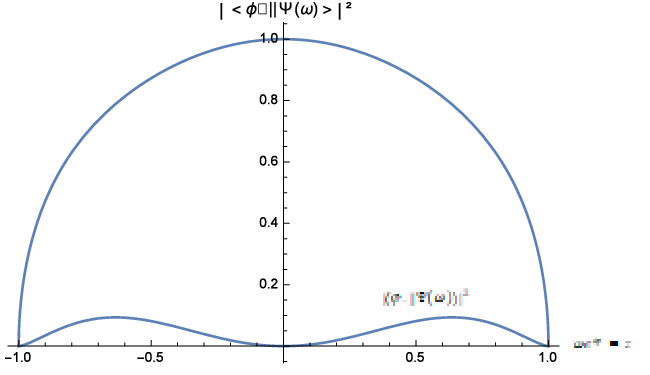
}
\caption{In the Figure we see graphically represented the square norm of the projections under the application of the Minimal Group Representation, against the variable $ z = \omega \, e^{\,i\, \varphi}$:  
 The huge curve  $\left\vert \,\left\langle\,
\varphi_{+}\right\vert \left\vert \,\Psi\left(  \omega\right)\,  \right\rangle\,
\right\vert ^{2}$ corresponds to the  $s = 1/4$, even $n$ sector in the Hilbert space of the
analytical functions, and  the small curve $\left\vert\, \left\langle \varphi_{-}\right\vert
\left\vert \Psi\left(  \omega\right)\,  \right\rangle\, \right\vert ^{2}$ corresponds to the  $s = 3/2$, odd $n$ sector in the Hilbert space of the analytical functions ). }%
\label{f4}%
\end{center}
\end{figure}

{\bf (3)} The analytic functions induced by the action of the basic states of the
metaplectic group and the London states divide the projection of the Hilbert
space on the disk into {\it even} and {\it odd} functions that in all cases (both for the
square of the norm and for the Wigner function) remain analytic inside the
unit disk, as well as for its analytic extension (by inversion). In the Figure
(\ref{f4}) we can clearly seen this fact.\\

{\bf (4)} The application of the MGRP to the coherent states of the quantum particle on
the cylinder (configuration space) of Refs \cite{kowalski1}, \cite{kowalski2}, \cite{gonzalez},
and Ref \cite {ciriloJPA} in the non-orientable  
case, {\it classicalizes the system} in a similar way to the London circle phase states but the
square norm of the scalar product between the $Mp(2)$ states and the cilinder configuration coherent
states has an extremely fast decay for increasing $n$ due to the content of the factors 
$e^{-2\,n^{2}}$,  $e^{-(\,2n+1/2\,)}$ and $e^{-(\,2n+1/2\,)^{2}}$, etc. \\
One of the reasons for the complexity
of the obtained functions is due to 
the construction of the coherent states on the circle such as the non-standard
modification of the Barut-Girardello definition for the Kowalski et al case,
or the introduction of a Gaussian fiducial state for the Ref \cite{gonzalez} proposal. \\

{\bf (5)} In order to elucidate the classical-quantum duality problem by considering
the circle topology in a complete way, {\it new coherent states for the circle} were
introduced here. These coherent states follow Perelomov's definition Ref. \cite{Perelomov} of coset coherent states, where the operators and the fiducial
vector are completely determined by the nonlinear realization of the coset of
the group E(2) on the group of translations in 2 dimensions as a stability
group: $g = E\left(  2\right)  /\mathbb{T}_{2}.$  These {\it new coherent states} solve
the identity but in a weak way (diagonal matrix with entries  $\mathbb{M}%
_{nn}\, = \;\, $e$^{-n\operatorname{Im}\alpha}\,)$,  and the most important thing is that
they are {\it completely normalizable}, well defined in the Hilbert space, contrary to
the London states. \\

From all the cases exhaustively studied here $ Mp(2)$ emerges as the {\it classical-quantum duality} group of symmetry .

\bigskip

\section{Acknowledgements}

\bigskip

DJCL acknowledges the institutions CONICET and the Keldysh Institute of Applied Mathematics, and the support of the Moscow Center of Fundamental and Applied Mathematics, Agreement with the Ministry of Science and Higher Education of the Russian Federation, No. 075-15-2022-283. NGS acknowledges useful communications with Gerard \\'t Hooft and José Luis Mac Loughlin on various occasions.


\bigskip

\begin{thebibliography} {}%

\bibitem{Nobel-2022} Alain Aspect, John F. Clauser and Anton Zeilinger, Nobel Prize in Physics 2022, "for experiments with entangled photons, ... and pioneering quantum information science" https://www.nobelprize.org/uploads/2023/10/advanced-physicsprize2022-4.pdf \\ and Refs therein. \\

\bibitem{MacLoughlin-Sanchez} J. L. Mac Loughlin, N. G. Sanchez,
"The Photography as a New Conceptual Quantum Information System", (Oct. 2023)\\
https://www.researchgate.net/publication/375745724 \\
https://www.bibsonomy.org/bibtex/7a77b52e38f27089f47229b1193dc005 \\

D. J. Cirilo-Lombardo and N. G. Sanchez, Entanglement and Generalized Berry Geometrical Phases in Quantum Gravity, 
{\it Symmetry}, 16(8), 1026 (2024) \\ https://doi.org/10.3390/sym16081026\\

\bibitem{Nobel-2024}John J. Hopfield and Geoffrey E. Hinton, Nobel Prize in Physics 2024, "for foundational discoveries and inventions that enable machine learning 
with artificial neural networks"
https://www.nobelprize.org/uploads/2024/10/advanced-physicsprize2024-2.pdf \\ 
and Refs therein. \\
 
\bibitem{tHooft1} G. 't Hooft, {\it The Hidden Ontological Variable in Quantum Harmonic Oscillators }, Lindau 2024 Lecture, arXiv 2407.18153, (August 2024).\\

\bibitem {london} F. London, {\it Winkelvariable und Kanonische Transformationen in der Undu-
lationsmechanik: Zeitschrift fur Physik}, 37, 915-925, 
 (1926).\\

\bibitem {universe} D. J. Cirilo-Lombardo, N. G. Sanchez, Universe, 10, 22
(2024).\\
https://doi.org/10.3390/universe10010022

\bibitem{cirilo-sanchezPRD} D. J. Cirilo-Lombardo, N. G. Sanchez, Phys. Rev.D108, 126001, (2023).\\

\bibitem {NSPRD2021} N. G. Sanchez, Phys. Rev. D 104, 123517 (2021).\\

\bibitem {NSPRD2023} N. G. Sanchez, Phys. Rev. D 107, 126018 (2023).\\

\bibitem {Sanchez2019a} N. G. Sanchez, Int. J. Mod Phys D28, 1950055 (2019).\\

\bibitem {Sanchez2019b} N. G. Sanchez, Int. J. Mod Phys A34, 1950155 (2019).\\

\bibitem {Sanchez2019c} N. G. Sanchez, 
 Gravit. Cosmol. 25, 91–102 (2019). \\ https://doi.org/10.1134/S0202289319020142 \\

\bibitem{pseu} D. J. Cirilo-Lombardo, J. Math.Phys. 57 (2016) 6, 063503\\

\bibitem{diego4} D. J. Cirilo-Lombardo; Foundations of Physics 37 (2007) 919-950.\\

\bibitem{diego5} D. J. Cirilo-Lombardo, Found Phys 39 (2009) 373--396.\\

\bibitem{Kla1} J. R. Klauder and B. S. Skagerstam, {\it Coherent States: Applications in
Physics and Mathematical Physics} (World Scientific, Singapore, 1985)\\

\bibitem{Kla} John R. Klauder and E. C. G. Sudarshan, \textit{Fundamentals
of Quantum Optics}, Dover Publications (2006)\\

\bibitem{pere} A. M. Perelomov, \textit{Generalized Coherent States and Their
Applications}, Springer Berlin, Heidelberg (1986).\\

\bibitem{kowalski1} K. Kowalski  et al  J. Phys. A: Math. Gen. 29 4149 (1996). \\

\bibitem{kowalski2} K. Kowalski and J. Rembielinski, Phys. Rev. A
75 052102, (2007). \\

\bibitem{gonzalez} J. A. Gonzalez and M. A. del Olmo, J. Phys. A: Math. Gen. 31 8841 (1998).\\

\bibitem{ciriloJPA} D. J. Cirilo-Lombardo, J. Phys. A: Math.
Theor. 45, 244026 (2012). \\

\bibitem{Perelomov} {\it Coherent
states for arbitrary Lie group}, A. M. Perelomov, Commun. Math. Phys. 26 (1972) 222--236].

\end{thebibliography}
\end{document}